\documentclass[conference]{IEEEtran}
\IEEEoverridecommandlockouts

\usepackage{cite}
\usepackage{amsmath,amssymb,amsfonts}
\usepackage{algorithmic}
\usepackage{graphicx}
\usepackage{textcomp}
\usepackage{xcolor}
\def\BibTeX{{\rm B\kern-.05em{\sc i\kern-.025em b}\kern-.08em
    T\kern-.1667em\lower.7ex\hbox{E}\kern-.125emX}}
\begin{document}
\title{Cognitive Skills in the Age of AI: Computing Students' and Experts' Perceptions
}
\author{\IEEEauthorblockN{1\textsuperscript{st} Neha Rani}
\IEEEauthorblockA{\textit{Computer Science} \\
\textit{University of Florida}\\
neharani@ufl.edu}
\and
\IEEEauthorblockN{2\textsuperscript{nd} Vu Minh Anh Le}
\IEEEauthorblockA{\textit{Computer Science} \\
\textit{University of Florida}\\
vu.le@ufl.edu}
\and
\IEEEauthorblockN{3\textsuperscript{rd} Austin M. Spangler}
\IEEEauthorblockA{\textit{Computer Science} \\
\textit{University of Florida}\\
aspangler1@ufl.edu}
\and
\IEEEauthorblockN{4\textsuperscript{th} Erta Cenko}
\IEEEauthorblockA{\textit{Epidemiology} \\
\textit{University of Florida}\\
ertacenko@ufl.edu}
}

\maketitle

\begin{abstract}
AI is becoming increasingly integrated into daily workflows, especially in computing. We are gradually shifting towards an AI-rich future, an impending yet unknown one. One important emerging concern is whether we are accordingly preparing our future computing workforce. Further, we need to know what the important cognitive skills are to remain relevant in the computing workforce and if there are changes in cognitive skill importance. To investigate this direction, we conducted a mixed-methods study, collecting perceptions from computing students and computing experts regarding the importance of cognitive skills in the past, present, and future. We report that the perceived importance of most cognitive skills will decrease in the future, with an AI-rich environment, but critical thinking skills remain important. Further, we report reasons collected through interviews on why the importance of cognitive skills will change and how future computing students can prepare for it.
\end{abstract} 


\section{Introduction}

Cognitive ability has long been regarded as a central objective of formal education. The American Psychological Association defines cognitive ability as the skills involved in performing the tasks associated with perception, learning, memory, understanding, awareness, reasoning, judgment, intuition, and language \cite{apa2018cognitiveability}. 
Researchers routinely use measures of cognitive ability to estimate academic success \cite{shi2022effect, deary2007intelligence}. Students are expected to develop and grow in such skills throughout their education and training, and apply them in professional work. Within computing disciplines in particular, students use cognitive skills such as critical thinking, working memory, problem solving, and more. 
In computing, these abilities have long underpinned hiring, particularly through technical interviews. \cite{bell2025software}. 
Educational institutions recognize that cognitive abilities such as reasoning, problem solving, and the flexible application of knowledge are widely recognized as core goals of formal education \cite{bransford2000how}.

The rapid and unprecedented diffusion of artificial intelligence into education introduces new uncertainty into this longstanding relationship between learning and cognitive skill development. 
A large-scale survey of over 1,000 full-time undergraduate students found that approximately 92\% had used generative AI tools in 2025 \cite{freeman2025student}. 
These findings suggest that AI use—whether formal or informal—has rapidly permeated educational environments. Students increasingly report using AI not only for routine tasks such as grammar correction, but also for higher-order academic activities requiring synthesis, analysis, and conceptual understanding \cite{black2025university}.

AI systems are increasingly capable of performing tasks traditionally associated with human cognition, including data analysis, logical reasoning, and knowledge retention \cite{korteling2021human}. As a result, AI systems can assist with many cognitive tasks that students historically performed independently. Critical thinking, a skill emphasized by various accreditation boards and educational boards for its educational necessity, may be influenced by the increasing availability of AI-assisted problem solving \cite{ahern2019literature}. 
Abstract thinking, systems thinking, and reasoning are also being offloaded to AI tools that can decompose problems and propose solutions. 

With growing pressures to prepare students for an AI-integrated workforce, understanding the impacts of AI on cognitive abilities and education is essential. We need to understand how cognitive skills will be reshaped in the emerging computing workplace. This will help prepare computing professionals for the AI-rich computing workspace. We conducted a mixed-methods study with 21 participants. In this study, we use surveys and interviews to investigate perceptions of AI’s impact on cognitive skills among undergraduate computing students and computing experts (instructors) at a large public research-intensive university in the United States. 


\section{Background and Related Work}

Prior research shows that instructors and students recognize the growing role of AI in computing education \cite{zastudil2023generativeai}. However, various authors have shared tensions between instructors' and students' perspectives. Instructors express concern about academic integrity and challenges in accurately assessing students' true abilities with the existence of AI. \cite{zastudil2023generativeai}. In contrast, students show greater excitement about learning with AI tools as personalized tutors, emphasizing improved efficiency and overcoming learning obstacles \cite{zastudil2023generativeai}. Other prior works also reinforce this divergence. 
Instructors' skepticism toward the growing use of AI in education, particularly concerns about overreliance on AI-generated content and the risk of misinformation in AI output \cite{verboom2025perceptionsai, pitts2026trust}, highlights their intention to preserve core cognitive skills. Research shows that early AI literacy may be critical, yet AI instruction is often introduced late in computing curricula \cite{MENDOZADIAZ2025100495}. 
Examining key cognitive skills in the AI age and perception gaps between students and instructors can inform curriculum design

The research questions we investigate in this study are.

{\textbf{RQ1: }What is the perceived importance of cognitive skills in the age of AI for computing professionals as perceived by engineering instructors? }

{\textbf{RQ2: }Is there a significant difference between the perceived importance of different cognitive skills between engineering students and instructors?}

{\textbf{RQ3: }What are the reasons for the difference in perception between engineering students and instructors regarding the importance of cognitive skills for computing professionals? }

\section{Study Design}
This was a researcher-guided one-to-one study session. The study used mixed methods and combined an interview and a survey. This IRB study took approximately 30 minutes. The study began with collecting participants' consent and then their demographics. 
Then, the researchers shared definitions of the different cognitive skills, laying the foundation for a shared understanding of the topic. A questionnaire was then provided, focusing on participants' perceptions of the importance of each cognitive skill for computing professionals. The same questionnaire was shared 3 times: past, present, and future, to gather their perceived importance of cognitive skills at the three times. For past, participants were asked to imagine it in 2015, when GenAI tools were not available. For future, participants were asked to imagine a future when AI becomes more embedded. A brief interview followed after each time (past, present, and future). 
The interview focused on whether their response to the importance of cognitive skills changed, and, if so, why and suggestions for preparing for future. 

The 11 cognitive skills assessed in the study were: Critical Thinking (The ability to evaluate information objectively and make reasoned judgments.); 
Problem Solving (The ability to find solutions to difficult or complex issues.); 
Decision-Making (The process of choosing the best course of action among alternatives.); 
Attention Control (The capacity to maintain focus on relevant stimuli and shift focus as needed.); 
Working Memory (The ability to hold and manipulate information over short periods.); 
Cognitive Flexibility (The ability to switch between different tasks or mental frameworks.); 
Abstract Reasoning (The ability to analyze information, detect patterns, and solve problems on a complex, conceptual level.); 
Metacognition (Awareness and regulation of one’s own thought processes.); Information Literacy (The ability to identify, evaluate, and use information effectively.); 
Creativity (The ability to produce novel ideas.); Systems Thinking (Understanding how parts of a system interact and influence one another.); 
Verbal reasoning (The ability to understand, analyze, and draw conclusions from written or spoken information.)

\section{Data Analysis}
There were 17 computer science students and 4 computer science experts as participants. Students were at the sophomore level or higher. The experts were computer science instructors. They had 5 to 18 years of teaching experience, with ages ranging from 32 to 43 and an average age of 36. 
Among the 17 students, 7 were female and 10 male; 7 were White, 9 Asian, and 1 African American. The average age was 19.8 years.
Among experts, there were 3 males and 1 female. 


We conducted a descriptive analysis of survey responses on the importance of cognitive skills over time. 
We conducted a repeated-measures ANOVA on the reported importance ratings of cognitive skills for instructors and students'. 
Further, we conducted a mixed ANOVA on self-reported importance ratings of cognitive skills, with expertise (student vs. expert) as the between-subjects factor and time (past, present, and future) as the within-subjects factor.

\begin{figure*}[h]
  \centering
  \includegraphics[width=1\textwidth]{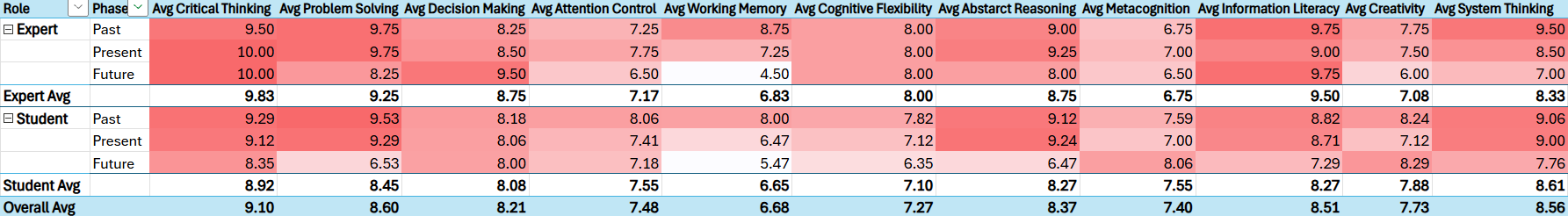}
  \caption{Average self‑reported importance ratings for cognitive skills, shown in a heatmap where darker tones indicate higher values.} 
  \label{All avg value}
\end{figure*}

\begin{figure*}[h]
  \centering
  \includegraphics[width=1\textwidth]{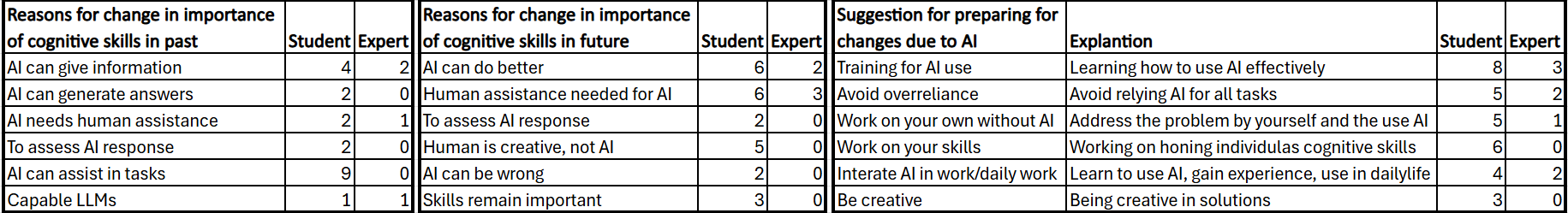}
  \caption{The frequency of different reasons for change in the importance of cognitive skills in the past and in the future is shown on the left. The table on the right shows suggestions for preparing for the future.} 
  \label{Reason for change in past and future}
\end{figure*}


We conducted a thematic analysis with intercoder reliability. We analyzed responses to the following interview questions: (i) Did your answers change (when rating importance from past to present)? Why or why not? (ii)Did your answers change (for future compared to past or present)? Why or why not? (iii)How do you think computing students or professionals should prepare for these changes?
All the interview audio was transcribed in an Excel sheet, separating responses for each question. Then, 3 researchers 
coded 10\% of the open-ended responses and set coding rubrics. Each response code and split into individual reasons for question (i) and (ii). Then the two coders split the data and coded the responses. 
Then coder 1 coded 30\% of the responses, which were coded by coder 2. The intercoder reliability was observed to be 57.1\% after the first round of coding. The response whether participants responded changed from present to future had a 100\% agreement. The coding for response to the reason for the change in importance had 52\% agreement. So both the coder and the reviewer reviewed 30\% of the response that both of them coded, discussed, and reached a common understanding. Both coders individually coded the entire dataset again. Then again, the researcher and the 2 coders met to determine intercoder reliability. After the second round of coding, intercoder reliability reached 85.7\%. 
The same process was followed for interview question (iii). Intercoder reliability was 71.5\%. 

\section{Results and Discussion}
The average scores for the importance of cognitive skills for students and experts are reported in Figure \ref{All avg value}. As shown in the figure, students reported the importance of different cognitive skills to decrease over time, i.e., from past to present to future. However, the importance of cognitive skills remains similar or reduces marginally from the expert perspectives. 
We observed a sharp decrease in the importance of working memory. Both experts and students believe that AI will be used to gather information as and when needed. 
Group sizes were unequal, with 4 experts and 17 students.



In response to whether they reported a change in the level of importance of cognitive skills from the present to the past. Of the 21 participants (4 experts and 17 students), 20 said they believe importance changes. Further, they reported that in current times, creativity, metacognition, systems thinking, and problem-solving are more important than in the past. While the importance of working memory, attention control, and problem-solving has decreased in the current times as compared to the past, when AI was less popular. 
We observed conflicting responses, some emphasized that problem‑solving remains essential for tackling complex tasks, while others felt people now rely on AI to handle much of the problem‑solving.
Participants explained that the importance of cognitive skills has shifted in current times, as AI now provides instant, human-like answers and can process natural-language prompts effectively. They noted that although AI is used more than in the past, it still requires human oversight, making skills like critical thinking important both then and now.
Many participants (8 of 14) felt that skills like working memory and problem‑solving will become less important as AI outperforms humans, while others argued the decline will be limited because people must still work with AI and evaluate its outputs. Views on creativity also conflicted: some believed it would become more important in an AI‑rich future, whereas others felt AI’s ability to generate creative ideas may reduce the need for human creativity.
Reasons for change in the importance of cognitive skills are shown in the left part of Figure \ref{Reason for change in past and future}.

The interview question (iii) focused on understanding from the viewpoints of both the student and the instructor.
All the coded suggestions to prepare for the future, with their explanations and frequencies, are reported in the right part of Figure \ref{Reason for change in past and future}.
The majority of participants (8 out of 17 students and 3 out of 4 instructors) said that preparing for an AI‑rich computing future requires stronger AI‑use training so students learn to use AI effectively and efficiently.
Many of them (7 out of 21) said that preparing for the future requires reducing AI overreliance rather than turning to AI for everything. 
For individuals to retain their skills, they should first solve the problem on their own, learn, and then use AI for assistance. In general, students should understand the importance of skills and focus on skill development. 
Some participants said that integrating AI into daily workflows helps prepare individuals for an AI‑rich environment. They also noted that creativity (generating original ideas and solutions) can distinguish humans from AI and help them remain relevant in the future.

To conclude, our findings show that the importance of major cognitive skills is expected to shift in an AI‑rich future. Students who are future computing professionals believe that the importance of most cognitive skills will decline, with critical thinking remaining essential, and working memory decreasing the most. Participants also offered suggestions for preparing for this future, 
provides directions for future research. 

\bibliographystyle{IEEEtran}
\bibliography{Reference}

\end{document}